\documentclass[%
 aip,
 amsmath,amssymb,
 preprint,%
]{revtex4-1}

\usepackage{lineno}

\usepackage{tikz}

\usepackage{hyperref}
\usepackage{url}

\usepackage[english]{babel}

\usepackage{aas_macros}

\usepackage{multirow}
\usepackage{physics}

\usepackage[symbol]{footmisc}

\usepackage{soul}

\usepackage[normalem]{ulem}

\usepackage{graphicx}
\usepackage{dcolumn}
\usepackage{bm}

\usepackage[utf8]{inputenc}
\usepackage[T1]{fontenc}
\usepackage{mathptmx}
\usepackage{etoolbox}

\makeatletter
\def\@email#1#2{%
 \endgroup
 \patchcmd{\titleblock@produce}
  {\frontmatter@RRAPformat}
  {\frontmatter@RRAPformat{\produce@RRAP{*#1\href{mailto:#2}{#2}}}\frontmatter@RRAPformat}
  {}{}
}%
\makeatother

\begin{document}

\preprint{ }

\title{Electromagnetic field solver for QED polarization in super-strong magnetic fields of magnetar and laser plasmas}

\author{M. Alawashra}
 \altaffiliation{M.~Alawashra and J.~Benáček contributed equally to this work.}
\affiliation{Deutsches Elektronen-Synchrotron DESY,\\ Platanenallee 6, 15738 Zeuthen, Germany}

\author{J. Benáček}
 \altaffiliation{M.~Alawashra and J.~Benáček contributed equally to this work.}
\affiliation{Institute for Physics and Astronomy, University of Potsdam,\\ Karl-Liebknech-Str. 24/25, 14476 Potsdam, Germany}

\author{M. Pohl}
\affiliation{Deutsches Elektronen-Synchrotron DESY,\\ Platanenallee 6, 15738 Zeuthen, Germany}
\affiliation{Institute for Physics and Astronomy, University of Potsdam,\\ Karl-Liebknech-Str. 24/25, 14476 Potsdam, Germany}

\author{and M.V. Medvedev}
\affiliation{Department of Physics and Astronomy, University of Kansas,\\ Lawrence, Kansas 66045, USA}
\affiliation{Laboratory for Nuclear Science, Massachusetts Institute of Technology,\\ Cambridge, Massachusetts 02139, USA}
 
\email{mahmoud.al-awashra@desy.de}
\email{jan.benacek@uni-potsdam.de}

\date{\today}

\begin{abstract}
Super-strongly magnetized plasmas play a crucial role in extreme environments of magnetar and laboratory laser experiments, demanding comprehensive understanding of how quantum electrodynamic (QED) effects influence plasma behaviour.
Earlier analytical and
semi-analytical calculations have shown that QED effects can significantly modify the plasma polarization mode behaviour around magnetars using analytical and semi-analytical calculations. In this work, we present the first electromagnetic
field solver that is valid beyond the Schwinger limit. QED vacuum polarization in super-strong
magnetic fields are modeled with nonlinear Maxwell equations.
We show that electromagnetic waves in simulations follow the analytical solutions well and reproduce the birefringence effects of electromagnetic wave modes between the $O$ and $X$ polarizations of perpendicular electromagnetic waves and those between $L$ and $R$ polarizations of parallel waves.
This new framework can be applied to kinetic as well as in other types of computer simulations.
The solver's key advantage lies in its versatility, allowing it to be used in gyro-motion, gyro-center, and gyro-kinetic simulations, which do not resolve the cyclotron motion, or in plasma studies with ground-level Landau quantization.
\end{abstract}

\maketitle
\flushbottom

\section{Introduction}
\label{sec:intro}

The magnetic field strength in magnetar magnetospheres as well as in laboratory laser plasma experiments can exceed the Schwinger quantum limit of $B_\mathrm{Q} \approx 4.4\times10^{13}$\,G, constraining the validity of classical theories of electromagnetism. In such super-strong magnetic fields, the vacuum behaves as a polarized medium because of the interaction of the electromagnetic fields with virtual electron--positron pairs. As a result, a photon travelling through this magnetized environment can undergo refraction or wave-splitting phenomena that are not allowed in classical electrodynamics.



The magnetars, often considered the ``central engines'' of gamma-ray bursts, X-ray flares, and fast radio bursts, have magnetic fields whose strength reaches $10^{15} - 10^{17}$\,G \cite{Mereghetti2005,Rowlinson2014,Younes2023}.
Some pulsars, such as PSR~J1846--0258 and PSR~J1119--6127, can also reach this limit, assuming they lose their rotational energy primarily through dipole radiation \cite{Gotthelf2000}.
For multipolar radiation, the magnetic fields of pulsars or magnetars are also predicted to exceed the quantum limit \cite{Gil2001}.
Therefore, highly magnetized neutron stars serve as natural laboratories for testing quantum electrodynamics (QED) in the strong-field regime, offering a unique opportunity to study its effects and implications.


Observations of magnetar X-ray polarization can provide a crucial test of the QED effects in super-strong magnetic fields. A key effect, vacuum birefringence, reflecting a modified vacuum dielectricity and permeability, can explain the X-ray polarization observed from magnetars like 4U~0142+61 \cite{doi:10.1126/science.add0080,Rigoselli2024,Taverna2024}.
The birefringence could provide photons at higher energies and those at lower energies with orthogonal polarization \cite{10.1093/mnras/stv2168}.


In the quantum regime of vacuum polarization, QED effects cause polarization-dependent refraction by modifying the wave speeds of the ordinary (O) and extraordinary (X) modes \cite{Schwinger1951,Thompson_2020,Perez-Garcia:2022kvz}. In the $O$ mode, the electric field oscillates in the plane of the propagation vector $\vb{k}$ and the magnetic field $\vb{B}$, while in the $X$ mode, it oscillates perpendicular to both. However, right- (R) and left- (L) circularly polarized modes, propagating parallel to $\vb{B}$, remain unaffected, preserving their classical behaviour. There are also several other effects on particle interaction in the quantum regime \cite{Kostenko2018,Kostenko2019}.

Previous studies on vacuum birefringence and quantum refraction in strong magnetic fields have relied on one-loop QED effective-action approaches \cite{PhysRevD.97.083001,Caiazzo_2019,10.1093/mnras/staa3428,10.1093/mnras/stac1571}. Such a formalism was used to study the vacuum birefringence and quantum refraction in pulsar magnetospheres, applying it to the Goldreich–Julian model \cite{Kim:2022fkt} and to a practical model of pulsar emission \cite{10.1093/mnras/stae1304}. Though insightful, these methods assume uniform fields and neglect nonlinear plasma dynamics. In contrast, kinetic simulations such as PIC simulations can provide a better picture by self-consistently solving nonlinear Maxwell’s equations including QED effects, capturing the nonlinear interactions and plasma feedback that is essential for describing realistic astrophysical and laboratory environments.

Recent laboratory experiments with lasers approach the quantum limit \cite{Zhang2020,Gonoskov2022}, in laser physics often denoted as the ``critical limit''.
The QED polarization correction for the intensity of electromagnetic fields below the Schwinger limit has been developed \cite{Grismayer2016b}, and pair-producing mechanisms have been added for laser plasma simulations \cite{Schoeffler2019,Grismayer_2021,Qu2022,Schoeffler2023}. The nonlinear Maxwell equations for the case of the weak fields ($E \ll E_\mathrm{Q}$ and $B \ll B_\mathrm{Q}$) were implemented in \cite{Grismayer_2021}, following the spatial and temporal evolution of the electromagnetic field in multi-dimensions, which can be suitable for the PIC loop. However, the numerical implementation for strong magnetic fields, $B \gg B_\mathrm{Q}$, remains open.



The weak-field approximation is valid only below the quantum field limit and therefore one needs to consider Maxwell's equations for super-strong magnetic fields to study plasma environments in the quantum regime. A general analytical form of the modified Maxwell's equation can be derived, if the electric field vanishes and the magnetic field is strong, $E \ll E_\mathrm{Q}$, $B \gtrsim B_\mathrm{Q}$,  \citep{2006physics...5038H,Lundin345665}. 

In this work, we introduce the first numerical algorithm for solving the spatial and temporal evolution of the electromagnetic field in the quantum regime. We develop a numeric electromagnetic-field solver that can be used in electromagnetic and kinetic plasma simulations of magnetar magnetospheres and laser plasmas.
Specifically, we aim to implement, employ, and test the field solver in the PIC simulation method.

The paper is structured as follows. 
In section \ref{sec:NLMW}, we introduce the QED polarization nonlinear effects in Maxwell's equations.
The new numerical scheme for field solver is developed and the PIC simulation setup is described in section~\ref{sec:method}.
The results of the numerical tests and stability are presented in section~\ref{sec:results}.
We discuss the strengths and limits of the proposed scheme in section~\ref{sec:discuss} and state conclusions in section~\ref{sec:conclude}.

\section{Nonlinear Maxwell’s equations}
\label{sec:NLMW}

In classical electrodynamics, the vacuum is considered an empty passive space in which particles and fields move and interact. Maxwell's equations, which describe these interactions, are linear in nature and are driven by charges and currents. However, when electromagnetic fields reach high intensities, quantum electrodynamics (QED) effects become significant, introducing nonlinearities into the Maxwell's equations. These nonlinearities are results of the polarization of the quantum vacuum.


The quantum vacuum fluctuations of virtual electron-positron pairs can mediate the exchange of energy and momentum between photons. In the effective field theory framework, the Heisenberg-Euler Lagrangian density, $\mathcal{L}_\mathrm{HE}$, encapsulates all orders of the one-loop photon-photon interaction processes. This Lagrangian adds as the leading correction to the standard Maxwell Lagrangian density, $\mathcal{L}_\mathrm{M}$, with the full Lagrangian density being $\mathcal{L} = \mathcal{L}_\mathrm{M} + \mathcal{L}_\mathrm{HE}$ where
\begin{equation}
    \mathcal{L}_\mathrm{M} = - \frac{1}{16 \pi} F_{\mu \nu}F^{\mu \nu} - \frac{1}{c}A_\mu j^\mu,
\end{equation}
and \cite{2006physics...5038H,Medvedev2023}
\begin{equation}\label{eq:HE}
    \mathcal{L}_\mathrm{HE} = \frac{m_\mathrm{e} c^2}{8\pi^2} \left( \frac{m_\mathrm{e} c}{\hbar} \right)^3 \int_0^\infty \frac{e^{-\eta}}{\eta^3} 
\left[ - \left( \eta a \cot \eta a \right) \left( \eta b \coth \eta b \right) + 1 - \frac{\eta^2}{3} (a^2 - b^2) \right] \mathrm{d}\eta.
\end{equation}

Here $F_{\mu \nu} = \partial_\mu A_\nu - \partial_\mu A_\nu$ is the electromagnetic field tensor, $A_\nu$ is the four-potential, $j^\mu$ is the four-current density and the parameters $a$ and $b$ are given by a covariant form
 \begin{equation}
     a = -\frac{i \hbar e}{\sqrt{2} m_\mathrm{e}^2 c^3} \left[ \left( \mathcal{F} + i \mathcal{G} \right)^{1/2} - \left( \mathcal{F} - i \mathcal{G} \right)^{1/2} \right],
 \end{equation}
 \begin{equation}
     b = \frac{\hbar e}{\sqrt{2} m_\mathrm{e}^2 c^3} \left[ \left( \mathcal{F} + i \mathcal{G} \right)^{1/2} + \left( \mathcal{F} - i \mathcal{G} \right)^{1/2} \right],
 \end{equation}
and therefore they are valid in all reference frame. Here the invariant quantities are
\begin{equation}
\mathcal{F} = \frac{1}{4} F_{\mu \nu} F^{\mu \nu} = \frac{1}{2} \left( \mathbf{B}^2 - \mathbf{E}^2 \right),
\end{equation}

\begin{equation}
\mathcal{G} = \frac{1}{4} \hat{F}_{\mu \nu} \hat{F}^{\mu \nu} = - \mathbf{B} \cdot \mathbf{E},
\end{equation}
where the Hodge dual tensor is given by $\hat{F}^{\mu \nu} = \frac{1}{2}\epsilon^{\mu \nu \gamma \delta}F_{\gamma \delta}$, where $\epsilon^{\mu \nu \gamma \delta}$ is the Levi–Civita symbol. The parameters $a$ and $b$ can be rewritten in the following form
\begin{equation}
    a = \frac{E}{E_\mathrm{Q}}, \qquad b = \frac{B}{B_\mathrm{Q}},
\end{equation}
where $E_\mathrm{Q}$ and $B_\mathrm{Q}$ are the quantum electric and magnetic field strengths. 

Though the general analytical form of the QED modified Maxwell equations is unknown, one can obtain an analytical form under different conditions. Modified Maxwell's equations can be obtained analytically in the case of electromagnetic fields smaller than the quantum field limit ($E \ll E_Q$ and $B \ll B_Q$). The nonlinear Maxwell's equations for the case of the weak fields were implemented in \cite{Grismayer_2021}. Here we consider Maxwell's equations in a super-strong magnetic field by following the general analytical form of the modified Maxwell's equations for a vanishing electric field, $E \ll E_\mathrm{Q}$, and arbitrarily strong magnetic field, $B \gtrsim B_\mathrm{Q}$,  \citep{2006physics...5038H,Lundin345665}. 

The electromagnetic field equations are obtained from the variational principle for action. By performing a variation of the fields and requiring the variation to vanish, $\delta S = \delta \int \mathrm{d}^4x\mathcal{L}$, we get the Euler-Lagrange equations
\begin{equation}
    \partial_{\mu} \frac{\partial \mathcal{L}}{\partial F_{\mu \nu}} = \frac{1}{2c} j^{\nu},
\end{equation}
substituting the full Lagrangian density including the Heisenberg-Euler Lagrangian density, equation \eqref{eq:HE}, the nonlinear Maxwell’s equations take the form \cite{Lundin345665}
\begin{equation}\label{eq:NLMW1}
    \gamma_{\mathcal{F}} \partial_{\mu} F^{\mu \nu} + \frac{1}{2} \left[ \gamma_{\mathcal{F} \mathcal{F}} F^{\mu \nu} F_{\alpha \beta} + \gamma_{\mathcal{G} \mathcal{G}} \hat{F}^{\mu \nu} \hat{F}_{\alpha \beta} + \gamma_{\mathcal{F} \mathcal{G}} \left( F^{\mu \nu} \hat{F}_{\alpha \beta} + \hat{F}^{\mu \nu} F_{\alpha \beta} \right) \right] \partial_{\mu} F^{\alpha \beta} = \frac{1}{c} j^{\nu},
\end{equation}

\begin{figure}
    \centering
    \includegraphics[width=0.6\textwidth]{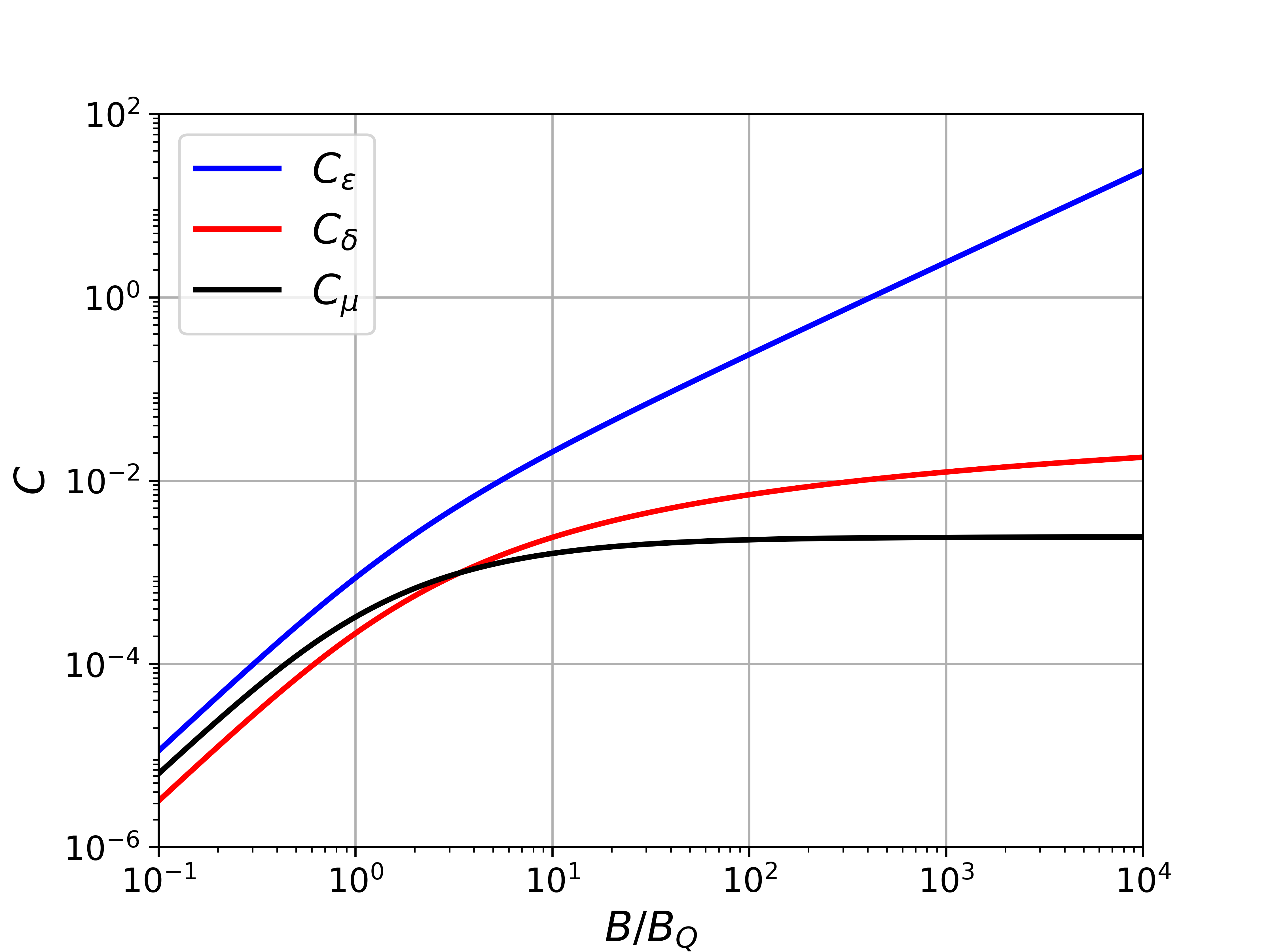}
    \caption{QED parameter $C_\epsilon$, $C_\delta$ and $C_\mu$ as functions of the normalized magnetic field $B/B_\mathrm{Q}$.
        \label{fig:C}
    }
\end{figure}

where QED coupling scalers are given as $\gamma_{\mathcal{F}} = \partial \mathcal{L}/ \partial \mathcal{F}$, $\gamma_{\mathcal{F} \mathcal{G}} = \partial^2 \mathcal{L}/ \partial \mathcal{F} \partial \mathcal{G}$, etc. For the case of vanishing electric fields, $\mathbf{E} \rightarrow 0$, and non-zero magnetic fields, $\mathbf{B} \neq 0$, the QED coupling scalers can be calculated analytically to be the following \cite{Lundin345665,Medvedev2023}
\begin{equation}\label{eq:C_delta}
\gamma_{\mathcal{F}} = \frac{-1+ C_\delta(b)}{4\pi},
\end{equation}
\begin{equation}
\gamma_{\mathcal{F} \mathcal{F}} = \frac{C_\mu(b)}{4\pi B^2},
\end{equation}
\begin{equation}\label{eq:C_epsilon}
\gamma_{\mathcal{G} \mathcal{G}} = \frac{C_\epsilon(b)}{4\pi B^2},
\end{equation}
\begin{equation}
\gamma_{\mathcal{F} \mathcal{G}} = 0,
\end{equation}
where the parameters $C_\delta$, $C_\mu$ and $C_\epsilon$ are all dependent on the normalized magnetic field, $b = B/B_Q$, and given analytically by the equations (28-30) in \cite{Medvedev2023}. Expressing equation \eqref{eq:NLMW1} explicitly in term of the electric, $\textbf{E}$, and magnetic, $\textbf{B}$, fields gives the modified Gauss' law
\begin{equation}
\begin{split}
\gamma_{\mathcal{F}} \, \nabla \cdot \mathbf{E} + \gamma_{\mathcal{F}\mathcal{F}} \, \mathbf{E} \cdot \nabla \left( \frac{\mathbf{B}^2 - \mathbf{E}^2}{2} \right) + \gamma_{\mathcal{G}\mathcal{G}} \, \mathbf{B} \cdot \nabla (-\mathbf{B} \cdot \mathbf{E})  = -\rho,
\end{split}
\end{equation}
and the modified Ampère's law
\begin{equation}\label{eq:ModAmpere}
\begin{split}
\gamma_{\mathcal{F}} & \left[ \frac{1}{c} \frac{\partial}{\partial t} \mathbf{E}  - \nabla \times \mathbf{B} \right] 
+ \gamma_{\mathcal{F}\mathcal{F}} \left[ \mathbf{E} \frac{1}{c} \frac{\partial}{\partial t} \left( \frac{\mathbf{B}^2 - \mathbf{E}^2}{2} \right) + \mathbf{B} \times \nabla \left( \frac{\mathbf{B}^2 - \mathbf{E}^2}{2} \right) \right] \\ &
+ \gamma_{\mathcal{G}\mathcal{G}} \left[ \mathbf{B} \frac{1}{c} \frac{\partial}{\partial t} (-\mathbf{B} \cdot \mathbf{E}) - \mathbf{E} \times \nabla (-\mathbf{B} \cdot \mathbf{E}) \right]  = \frac{1}{c} \mathbf{j}.
\end{split}
\end{equation}
As for the other pair of Maxwell's equations, $\partial_\nu \hat{F}^{\mu \nu} =0$, we retain the same equations as the classical case
\begin{equation}
    \nabla \cdot \mathbf{B} = 0,
\end{equation}
and 
\begin{equation}\label{eq:Faraday}
    \frac{1}{c} \frac{\partial \mathbf{B}}{\partial t} + \nabla \times \mathbf{E} = 0.
\end{equation}
Therefore, including the QED corrections in PIC simulation, we expect a modification of the electric field advance (Ampère's law) and no modification of the magnetic field one.

\section{Numerical methods}
\label{sec:method}

In this section, we introduce the main features of the new PIC field solver including the QED polarization effects. The standard PIC method uses a set of macroparticles with fixed shapes to represent the distribution function. This approach allows the distribution function to be updated by tracking the movement of the computational particles, while the Maxwell equations are solved on a discretized spatial grid. First, we introduce a novel field solver in section~\ref{sec:EMsolver} that solves the nonlinear Maxwell equations in the presence of super-strong magnetic fields. Then, we present the implementation of this solver in the PIC algorithm in section~\ref{sec:PIC}. Numerical effects arising from space and time discretization are briefly discussed in section~\ref{sec:NS}. Finally, in section~\ref{sec:SS} we present the simulation setup we used to test the new field solver. More details are given in appendices~\ref{app:EMsolver} and \ref{app:NS}.

\subsection{Electromagnetic solver in super-strong magnetic fields approach}\label{sec:EMsolver}

The widely used second-order finite-difference time-domain (FDTD) method for solving Maxwell’s equations on a spatial grid is the Yee scheme \cite{Yee1966}. This scheme simultaneously computes the magnetic and electric fields by discretizing and solving Faraday’s and Ampère’s laws, respectively. The Yee scheme uses a leapfrog approach, where fields are shifted in space and time, thus ensuring robustness and second-order accuracy while eliminating the need to solve coupled equations or perform matrix inversions. This leapfrog approach is only possible due to the explicit linear relationship between the fields in Maxwell's equations. Furthermore, the algorithm’s simplicity and computational efficiency make it well-suited for implementation in parallelized numerical codes, facilitating the performance of large-scale simulations.

We developed a modified Yee electric field solver on the Yee lattice that uses the QED-modified Ampère law in the strong-field limit (equation \eqref{eq:ModAmpere}). We see in equation \eqref{eq:ModAmpere} that the temporal derivatives of the different electric field components are mixed in each spatial component of the equation, unlike the linear relation in Ampère law. Therefore, we need first to isolate the temporal derivatives of each electric field component in one equation. We have outlined a series of  analytical procedures to achieve this in the appendix \ref{app:EMsolver}, finding the following expression for the temporal derivative of the electric field
\begin{equation}\label{eq:dEdt}
    \frac{\partial \mathbf{E}}{\partial t} = A^{-1} \left( \frac{1}{c} \vb{j} - \vb{Q}\right),
\end{equation}
where $A^{-1}$ is the inverse matrix of an $3 \times 3$ matrix $A$ given by 
\begin{equation}
    A_{ij} = \frac{1}{c} \left[ \gamma_{\mathcal{F}} \delta_{ij} - \gamma_{\mathcal{F} \mathcal{F}} E_i E_j - \gamma_{\mathcal{G} \mathcal{G}} B_i B_j \right],
\end{equation}
with the indices $i$ and $j$ both looping over the spatial dimensions $x$, $y$ and $z$, $\vb{j}$ is the electric current vector and $\vb{Q}$ is a vector with dependence on the spatial derivatives of the electric and magnetic fields and the magnetic field temporal derivative. The components of the vector, $\vb{Q}$, are given by equations \eqref{eq:Qx} - \eqref{eq:Qz}. For an accurate estimate of the magnetic field time derivative in $\vb{Q}$ (also seen in equation~\eqref{eq:ModAmpere}), we substitute its value from Faraday's law (equation \eqref{eq:Faraday}), yielding at the end the vector $\vb{Q}$ to be dependent only on the spatial derivatives of the electromagnetic fields. The numerical implementation of this field solver (equation \eqref{eq:dEdt}) is available at the Zenodo platform (DOI: \url{https://doi.org/10.5281/zenodo.15004304}).

\subsection{Updated Particle-in-cell method} \label{sec:PIC}

\begin{figure}
    \centering
    \includegraphics[width=\textwidth]{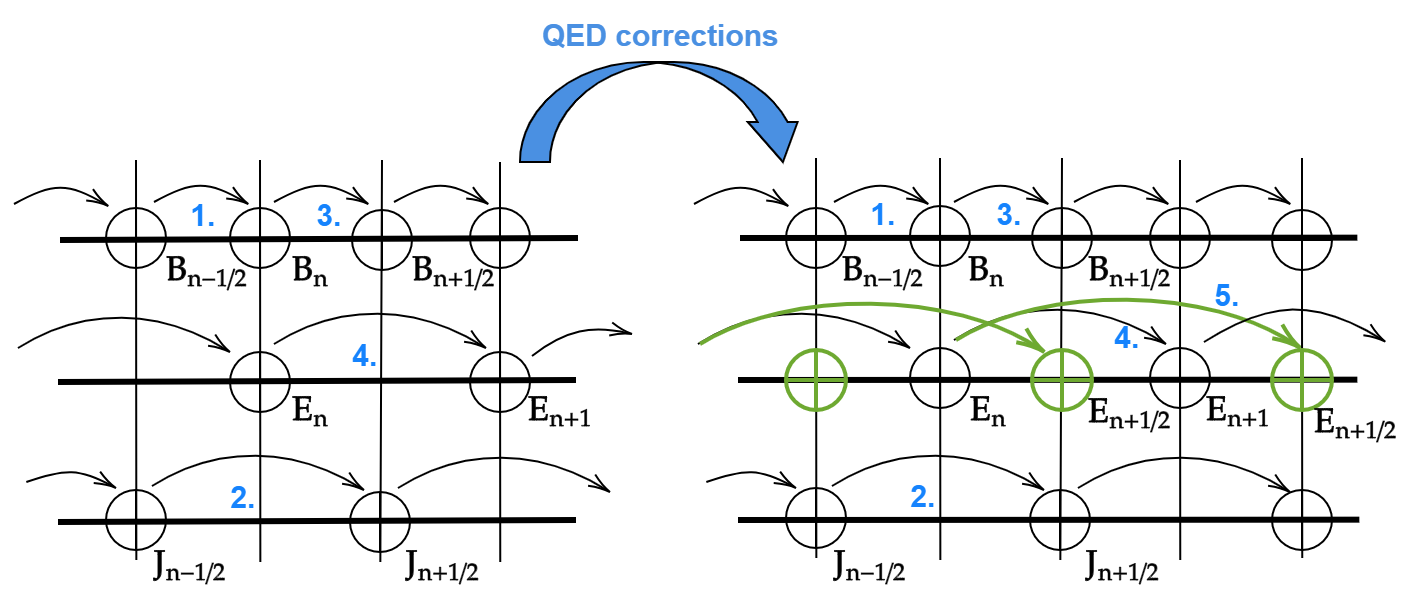} 
    \caption{The original (left) and QED polarization updated (right) PIC scheme for the advancement of plasma quantities.
    The new steps 4 and 5 are time-centered at timestep $n + 1/2$ and employ the usage of the new solver.}
    \label{fig:ampere-mods}
\end{figure}

We have updated the PIC algorithm
by the nonlinearity of the modified Ampère equation \eqref{eq:dEdt}. That requires additional steps in the leapfrog algorithm. Here we summarize those modifications. 

Note that the nonlinear Ampère law
given in equation \ref{eq:dEdt}, couples all the field components of $\boldsymbol{E}$ and $\boldsymbol{B}$. Therefore, the numerical solution of equation \eqref{eq:dEdt} requires the evaluation of the fields at all the grid points where the electric field is shifted on the Yee lattice. To do this, we linearly interpolate the fields using the nearest grid points.

We also note that solving equation \eqref{eq:dEdt} with the centered-difference time partial derivative requires knowledge of $\boldsymbol{E}$ at the half-time step. To preserve the same order of the time derivative of the Yee scheme, we first predict the electric fields at the half-time step using the QED modified solver in forward time derivatives and then we use the fields at the half-time step to evolve the electric field full timestep using centered time derivative. 
The comparison of the original and the new method is shown in figure~\ref{fig:ampere-mods}.

\subsection{Numerical Stability}\label{sec:NS}

The Yee algorithm can produce nonphysical effects such as numerical wave dispersion, when applied to nonlinear or linear Maxwell's equations. This is due to the finite differencing of spatial and temporal derivatives of the Yee scheme.
For example, the phase velocity of numerical wave modes may deviate from the speed of light c, depending on the wavelength and propagation direction. This discrepancy results in phase errors or delays of the propagating waves, leading to artefacts. Therefore, understanding the numerical dispersion is essential for assessing the Yee algorithm's behaviour and accuracy limits, particularly for the cases when the super-strong magnetic fields change linear Maxwell's equations significantly. 

We employ the standard mode analysis approach to analyze the numerical stability of the new QED polarization solver \citep{GREENWOOD2004665}. In the case of linear Maxwell's equation, the plasma dispersion relation for a plane electromagnetic wave propagating in a one-dimensional Yee grid is
\begin{equation}\label{eq:dis_c}
    \left( \frac{c \Delta t}{\Delta x} \right)^2 \sin^2 \left( \frac{k \Delta x}{2} \right) - \sin^2 \left( \frac{\omega \Delta t}{2} \right) = 0,
\end{equation}
where $\Delta x$ and $\Delta t$ are the spatial and time steps, respectively, and $\omega$ and $k$ are the wave frequency and wave number, respectively. Notice that for the case, $\Delta t = \Delta x/c$, the numerical dispersion relation recovers the physical electromagnetic dispersion relation, $\omega = kc$.

In appendix \ref{app:NS}, we derive the numerical dispersion relation for the nonlinear Maxwell's equations in the super-strong magnetic field case. We introduce here the final result of the dispersion relations and one can refer to the detailed calculations in the appendix. 

We consider a plane wave propagating in the $x$ direction and derive the dispersion relations for a homogeneous magnetic field in the $x-y$ plane, $\mathbf{B_0} = \text{B}_0 \cos{\theta} \hat{x} + \text{B}_0\sin{\theta} \hat{y}$. For the extraordinary electromagnetic mode ($X$ mode) and $\theta \neq 0$, where the wave-like perturbation components are $\delta B_\mathrm{y}$, $\delta E_\mathrm{x}$ and $\delta E_\mathrm{z}$, we found the following dispersion relation (equation~\eqref{eq:Xmode})
\begin{equation} \label{eq:dis_O}
    \left( \frac{1-C_\delta - C_\mu \sin^2{\theta}}{1-C_\delta} \right)\left( \frac{c \Delta t}{\Delta x} \right)^2 \sin^2 \left( \frac{k \Delta x}{2} \right) - \sin^2 \left( \frac{\omega \Delta t}{2} \right) = 0,
\end{equation}
where $C_\delta$ and $C_\mu$ are the QED parameters given in figure~\ref{fig:C} as a function of the homogeneous magnetic field strength.

\begin{figure}
    \centering
    \begin{minipage}[b]{0.47\textwidth}
        \centering
        \includegraphics[width=\textwidth]{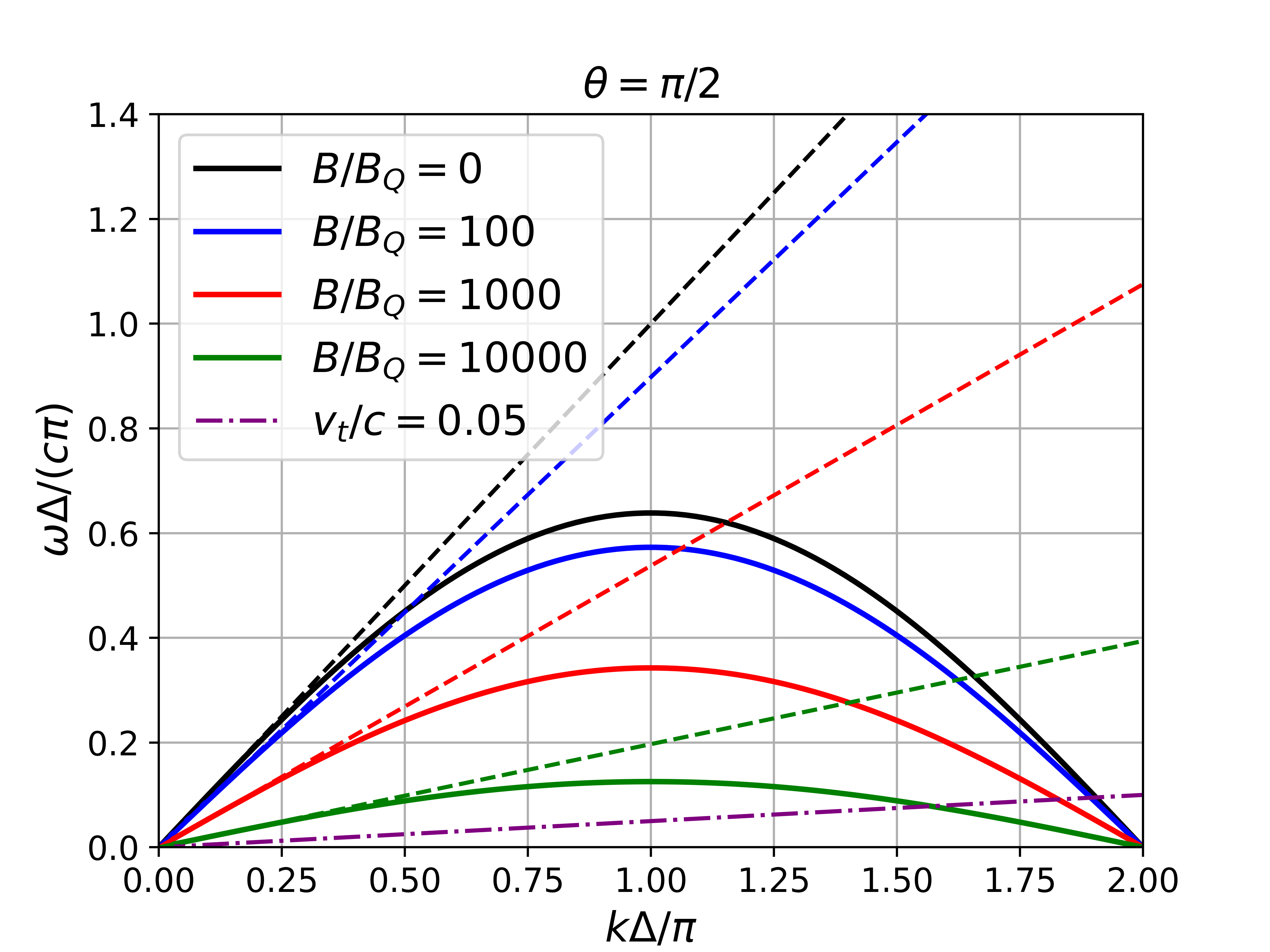} 
    \end{minipage}
    \hfill
    \begin{minipage}[b]{0.47\textwidth}
        \centering
        \includegraphics[width=\textwidth]{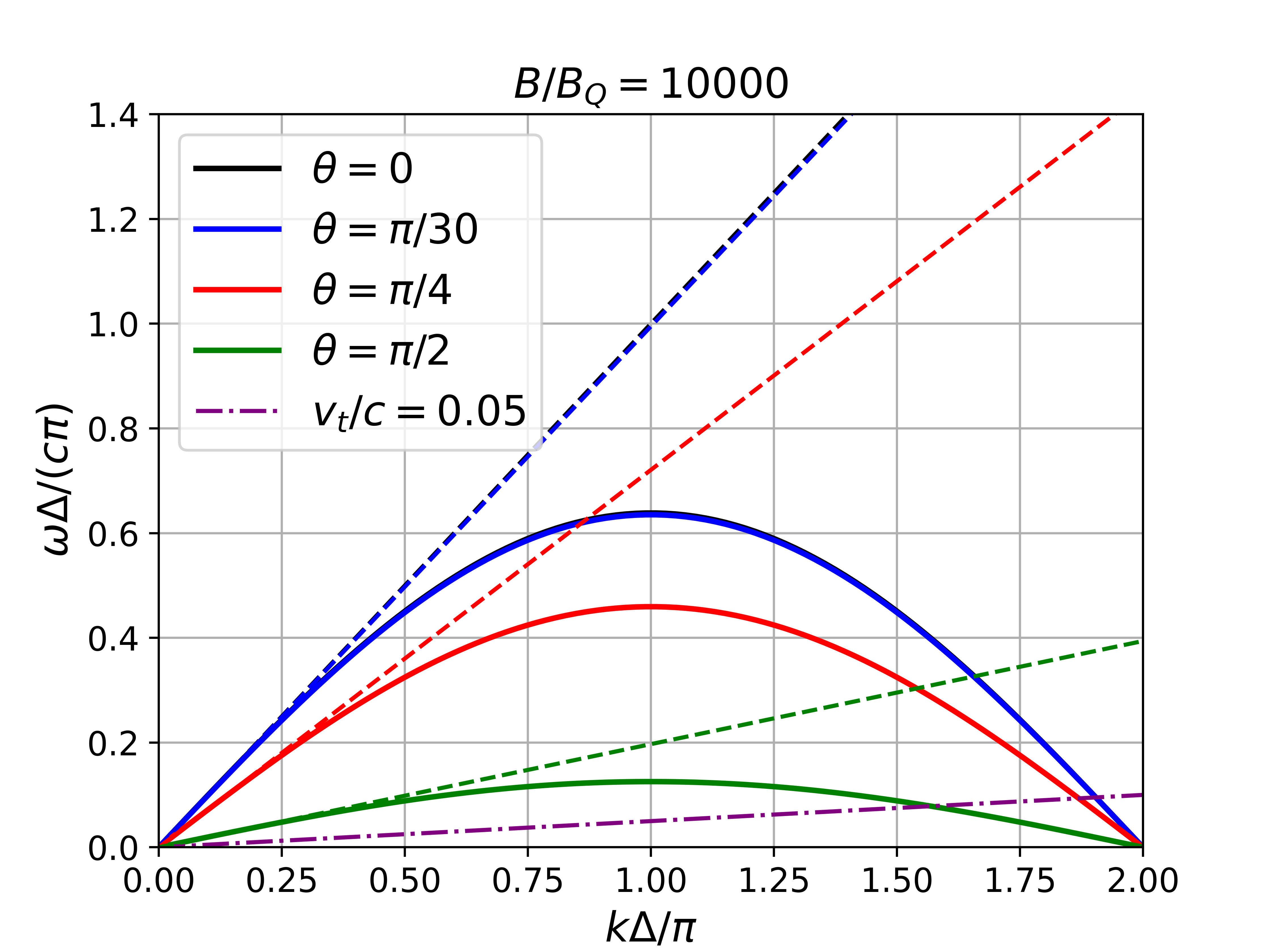} 
    \end{minipage}
    \caption{Dispersion relation of the O-mode waves including the QED modification, given by equation \ref{eq:dis_O}. Solid lines are the numerical dispersion curves, dashed lines are the analytically derived dispersion curves. The left figure includes the waves propagating perpendicular, $\theta = \pi/2$, to the homogeneous magnetic field with different strengths, $B/B_Q$. The right figure includes the waves propagating with different obliquity angles to the homogeneous magnetic fields with strength $B/B_Q = 10000$. The purple dashed-dotted curve shows the slope of the thermal velocity of the particles in the test PIC simulation in section \ref{sec:SS}. The curves with $B/B_Q = 0$ and $\theta =0 $ are identical. The FDTD parameters are $\Delta_x = \Delta$ and $\Delta t = 0.4 \Delta /c$.}
    \label{fig:Omode}
\end{figure}

The modification of the $X$ mode dispersion relation in equation~\eqref{eq:dis_O} compared to the classical one in equation~\eqref{eq:dis_c}, depends on $C_\mu \sin^2{\theta}$. Recall from figure~\ref{fig:C} that the value of the parameter $C_\mu$ saturates at a value much less than $10^{-2}$. This means there is no effect on the $X$ mode waves propagating parallel to the homogeneous magnetic field, $\theta =0$, and a negligible modification that is much less than one percent for the ones propagating perpendicular to the homogeneous magnetic field. 

For the Ordinary electromagnetic mode ($O$ mode), where the wave-like perturbation components are $\delta B_\mathrm{z}$, $\delta E_\mathrm{x}$ and $\delta E_\mathrm{y}$, the dispersion relation (equation~\eqref{eq:Omode}) gets the form
\begin{equation} \label{eq:dis_1}
    \left( \frac{1-C_\delta + C_\epsilon \cos^2{\theta}}{1-C_\delta + C_\epsilon} \right)\left( \frac{c \Delta t}{\Delta x} \right)^2 \sin^2 \left( \frac{k \Delta x}{2} \right) - \sin^2 \left( \frac{\omega \Delta t}{2} \right) = 0,
\end{equation}
where $C_\delta$ and $C_\epsilon$ are the QED parameters given in figure~\ref{fig:C} as a function of the homogeneous magnetic field strength. The QED modification of the $O$ mode dispersion relation is significant compared to the $O$ mode; in figure~\ref{fig:Omode}, we see this as a severe reduction of the phase speed of the O-mode waves.
Both the physical and the numerical dispersion phase speeds of the O-mode waves decrease as the homogeneous magnetic field increases and as the angle between the magnetic field and the wave vector approaches $\frac{\pi}{2}$. The numerical speed is always less than the physical light speed for the same parameters, which ensures the stability of the simulation.

Note that the deviation between physical and numerical dispersion figure \ref{fig:Omode} has nothing to do with the QED modification, as the same discrepancy between the two curves manifests for the weak field regime as well. We also see in figure \ref{fig:Omode} that the thermal speed of the particles in the performed test simulations in the section \ref{sec:results} is always less than the phase speed at short wave numbers. Therefore, we don't expect any Cherenkov radiation at short wave numbers in those simulations. However, as the magnetars' magnetospheric plasma is relativistic, the expected amount of the numerical Cherenkov radiation can be higher in simulations with relativistic setups.

\subsection{Simulation Setup}\label{sec:SS}

\begin{table}
\centering
\begin{tabular}{l|c}
\hline
Parameter & Value \\
\hline
Magnetic field intensities $B/B_\mathrm{Q}$ & 100, 1000, and 10\,000 \\
Magnetic field angles $\theta$ & $\pi/30$, $\pi/4$, and $\pi/2$ \\
Frequency ratio $\omega_\mathrm{c}/\omega_\mathrm{p}$ & 3 \\
Initial thermal velocity $v_\mathrm{t}/c$ & 0.05 \\
Simulation length $L / \Delta$ & 20\,000 \\
Simulation time $\omega_\mathrm{p} t_\mathrm{end}$ & 800 \\
Skin depth resolution $\Delta / d_\mathrm{e}$ & 0.05 \\
Time step size $\omega_\mathrm{p} \Delta t$ & 0.02 \\
\hline
\end{tabular}
\caption{
    Summary of used plasma and simulation parameters.
    \label{tab:1}}
\end{table}

To demonstrate the proposed field solver functionality, we implemented the numerical scheme of QED polarization into 1D3V (one spatial dimension, three velocity components) version of our code ACRONYM \cite{Kilian2012} on the rectangular Yee grid \cite{Yee1966}.
Though ACRONYM generally has a few high-order field solvers, our implementation is second-order.
However, we utilize a high-order current-conserving deposition scheme with a fourth-order piecewise quadratic shape (PQS) function for macro-particles \cite{Esikperov2001}.
We use the Vay \cite{Vay2011} particle pusher.
No further modifications of the particle dynamics in the quantum regime are implemented.
The simulation macro-particles interact with the electromagnetic fields using the classical Lorentz force.

The summary of plasma and simulation parameters is given in table~\ref{tab:1}.
We carry out a series of simulations of thermal plasma in strong external magnetic fields of intensities $B/B_\mathrm{Q} = 100$, 1000, and 10\,000.
In all cases, the plasma consists of electrons and positrons, and it has initially a uniform density $n$ and is composed of a certain ratio of electrons and positrons. The used non-neutral fractions of the electrons and positrons plasma $\Delta n / n = (n_\mathrm{p} - n_\mathrm{e})/n$ are 0, 0.5, and 1, where $n_\mathrm{e}$ is the electron density, $n_\mathrm{p}$ is the positron density and $n= n_\mathrm{e} + n_\mathrm{p}$ is the total plasma density.

The specific value of plasma density is obtained from the cyclotron to plasma frequency ratio, which we fix to a value of $\omega_\mathrm{c}/\omega_\mathrm{p} = 3$, where $\omega_\mathrm{c} = e B / m_\mathrm{e}$ is the electron cyclotron frequency.
The plasma density is in all cases represented by 40 macro-particles per cell.
The used ratio $\omega_\mathrm{c}/\omega_\mathrm{p} = 3$ is unrealistically low, producing plasma densities with particles closer to each other than the de~Broglie wavelength.
Nonetheless, we use this value for test purposes because the ratio (1) separates the cyclotron and plasma frequency effects sufficiently to resolve the plasma dispersion properties at both frequencies and (2) it reduces computational requirements because of the necessity to resolve the cyclotron motion as well as sufficiently long time in terms of the plasma periods.

The initial thermal velocity is $v_\mathrm{t}/c = 0.05$ with Maxwellian velocity distribution.
We consider three angles $\theta = \pi/30, \pi/4$, and $\pi/2$ between the magnetic field vector and the simulation domain $x$-axis to study quasi-parallel, oblique, and perpendicular wave types.
Also, the magnetic field intensities $B/B_\mathrm{Q}$ and propagation angles $\theta$ are selected for a comparison of analytical dispersion properties obtained in \cite{Medvedev2023}.

The simulation domain is $L = 20\,000\Delta = 1000 d_\mathrm{e}$ long, where $\Delta = 0.05 d_\mathrm{e}$ is the grid cell size and $d_\mathrm{e} = c / \omega_\mathrm{p}$ is the plasma skin depth.
The wavenumber resolution is $k c  / \omega_\mathrm{p} = 6.3\times 10^{-3}$.
The simulation time step is $\Delta t = 0.02\omega_\mathrm{p}^{-1} = 0.06\omega_\mathrm{c}^{-1}$.
Therefore, the stability conditions of equations~\eqref{eq:dis_O} and \eqref{eq:dis_1} are fulfilled for all simulation setups.
The simulation is carried out for 40\,000$\Delta t$, i.e., $\omega_\mathrm{p} t_\mathrm{end} = 800$, allowing frequency resolution of $\Delta  \omega / \omega_\mathrm{p} = 7.9\times 10^{-3}$.
We use periodic boundary conditions.

\section{Results}
\label{sec:results}

\begin{figure}
    \centering
    \includegraphics[width=\textwidth]{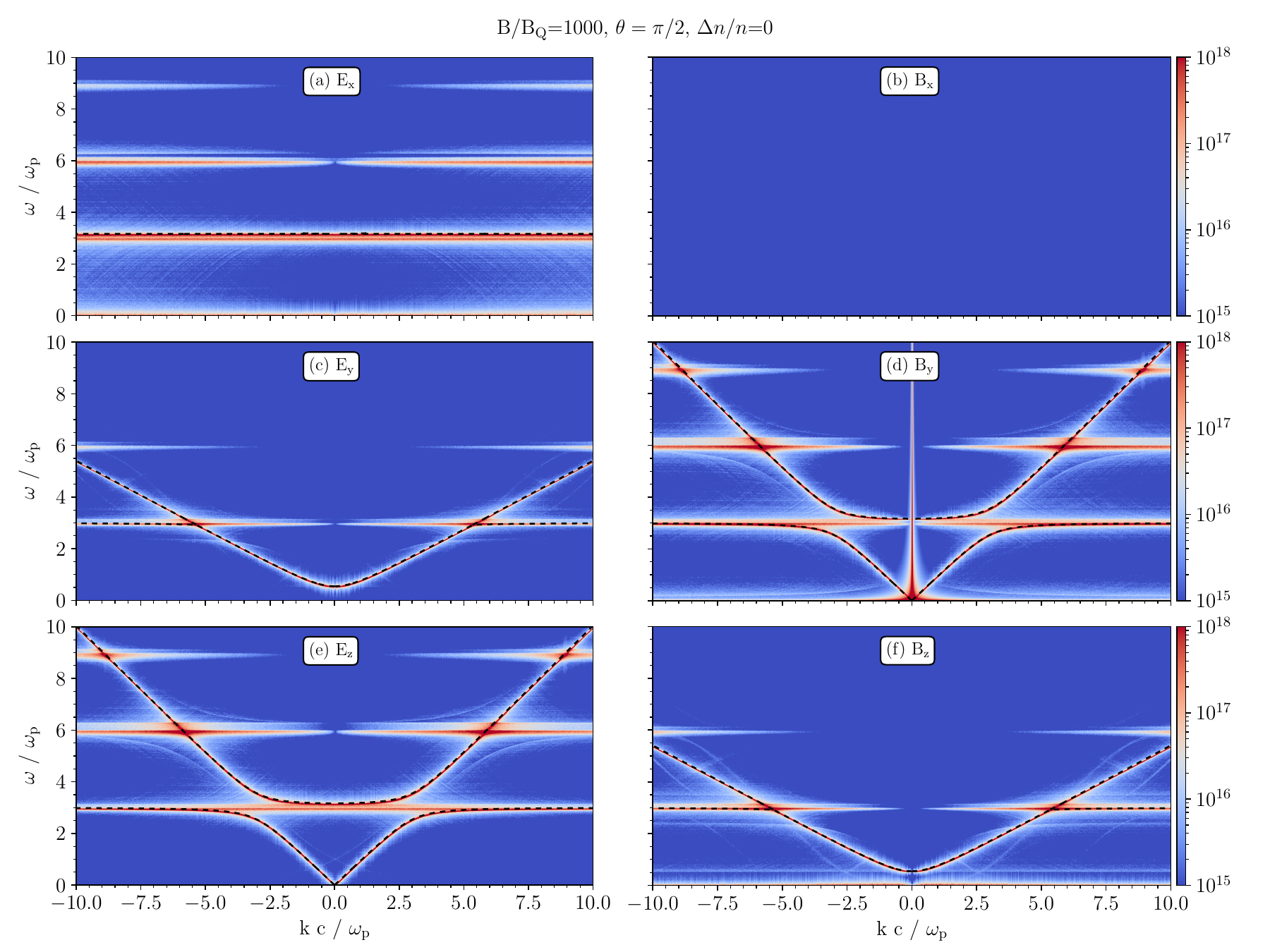}
    \caption{
        The dispersion diagrams of the electric and magnetic field components show the wave birefringence between the $O$ ($\text{E}_\text{y}$ and $\text{B}_\text{z}$ components) and $X$ mode ($\text{E}_\text{z}$ and $\text{B}_\text{y}$ components) waves for $B/B_\mathrm{Q} = 1000$, $\theta = \pi / 2$, and $\Delta n / n = 0$.
        \label{fig1}
    }
\end{figure}

The results of tests of plasma dispersion properties in super-strong magnetic fields are presented in figures~\ref{fig1}--\ref{fig4}.
The figures are overlaid by black dashed lines representing the solution of linearized dispersion solutions for the cold plasma approach obtained in \cite{Medvedev2023}.
In all the studied cases, the total energy in the simulation is conserved well with oscillation in approximately the same range as in a separate test simulation without the QED effects.

\subsection{Plasma dispersion properties with QED polarization effects}
Figure~\ref{fig1} shows the dispersion space of electric and magnetic field components for the case of $B/B_\mathrm{Q}=1000$ in the perpendicular direction to the magnetic field with a density ratio equal to one.
The perpendicular direction manifests the birefringence effect between $O$ and $X$ modes.
The $E_\mathrm{x}$ component shows the electrostatic cyclotron waves propagating perpendicular to the magnetic field and appearing mostly at the cyclotron frequency $\omega_\mathrm{c} = 3\omega_\mathrm{p}$ and its harmonics $2\omega_\mathrm{c}$ and $3\omega_\mathrm{c}$.
The $B_\mathrm{x}$ component is zero throughout the simulation domain as expected because $\partial B_\mathrm{x} / \partial t = \partial_\mathrm{y} B_\mathrm{z} - \partial_\mathrm{z} B_\mathrm{y} = 0$ in 1D since $\partial_\mathrm{y} f(x) = \partial_\mathrm{z} f(x) = 0$ for an arbitrary function $f(x)$.

The dispersion branches in $E_\mathrm{y}$ and $B_\mathrm{z}$ represent the $O$ mode waves and $E_\mathrm{z}$ and $B_\mathrm{y}$ represent the $X$ mode waves. 
Each combination of electric and magnetic field components corresponds to the same mode.
The frequency cut-off $\omega_\mathrm{co}$ is the same for both components for each polarization mode.
The main difference occurs close to the ``light line,'' $\omega=kc$, for $(\omega,k) \rightarrow \infty$.
When QED polarization is considered, the refractive index $N^2 = k^2 c^2 / \omega^2 = c^2 / v_\phi^2$ of the $O$ mode waves increases, where $v_\phi$ is the phase velocity.
The phase and group velocities of waves $(\omega,k) \rightarrow \infty$ decrease.
For the $X$ mode waves, $N \approx 1$ for $(\omega,k) \rightarrow \infty$ creates the birefringence between the $O$ and $X$ modes.
The dispersion properties of the $X$ mode waves do not change significantly from the non-QED approach.

The electromagnetic waves manifest increased intensity at the harmonics of the cyclotron frequency.
For the harmonics that appear for the electrostatic as well as electromagnetic waves and frequencies $\omega > \omega_\mathrm{c}$, there are no analytical solutions because they result from nonlinear wave--wave and particle--wave interactions, which are not described by the linearized analytical approach.

\subsection{Tests for broad plasma parameter space}

\begin{figure}
    \centering
    \includegraphics[width=\textwidth]{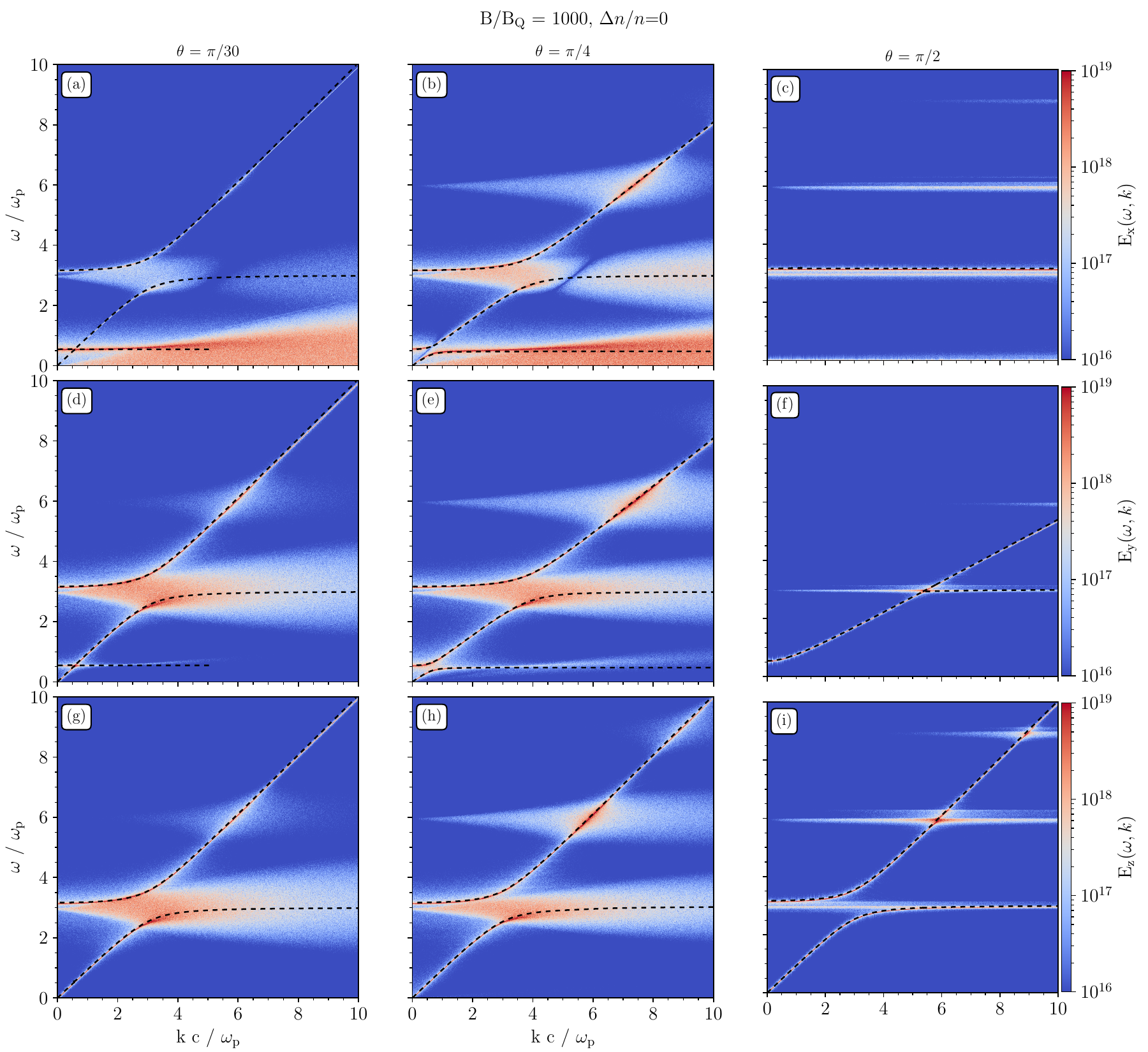}
    \caption{
        The dispersion diagrams of the electric field components for changing propagation angle to the magnetic field $\theta$ for $B/B_\mathrm{Q} = 1000$ and $\Delta n / n = 0$.
        \label{fig2}
    }
\end{figure}

The electric field dispersion properties as a function of $\theta$ are shown in figure~\ref{fig2} for an electron/positron density ratio of unity. We do not present the magnetic field components because they show the same.
The dispersion space of the $E_\mathrm{x}$ component for $\theta = \pi/30$ consists mostly of quasi-parallel electrostatic Langmuir waves close to the plasma frequency; however, the electromagnetic branches are also projected into $E_\mathrm{x}$.
The electrostatic mode close to $\omega_\mathrm{p}$ bends up with increasing $k$ while the analytical solution does not because, in comparison with the analytical solution obtained for cold plasma, we consider thermal plasma.
The Langmuir branch, which is smooth for parallel propagation, separates for an increased angle $\theta = \pi/4$ into two branches close to $\omega / \omega_\mathrm{p} \approx \ k c / \omega_\mathrm{p} \approx 0.6$.
The resulting branches have superluminal and subluminal parts that do not intersect the light line $\omega = kc$.
Only the electrostatic cyclotron mode appears for perpendicular waves at $\theta = \pi/2$.
The $E_\mathrm{y}$ component representing the $O$ mode waves undergoes a decrease in the wave phase and group velocities with an increasing $\theta$.
The $X$ mode properties of $E_\mathrm{z}$ component do not significantly depend on $\theta$.

\begin{figure}
    \centering
    \includegraphics[width=\textwidth]{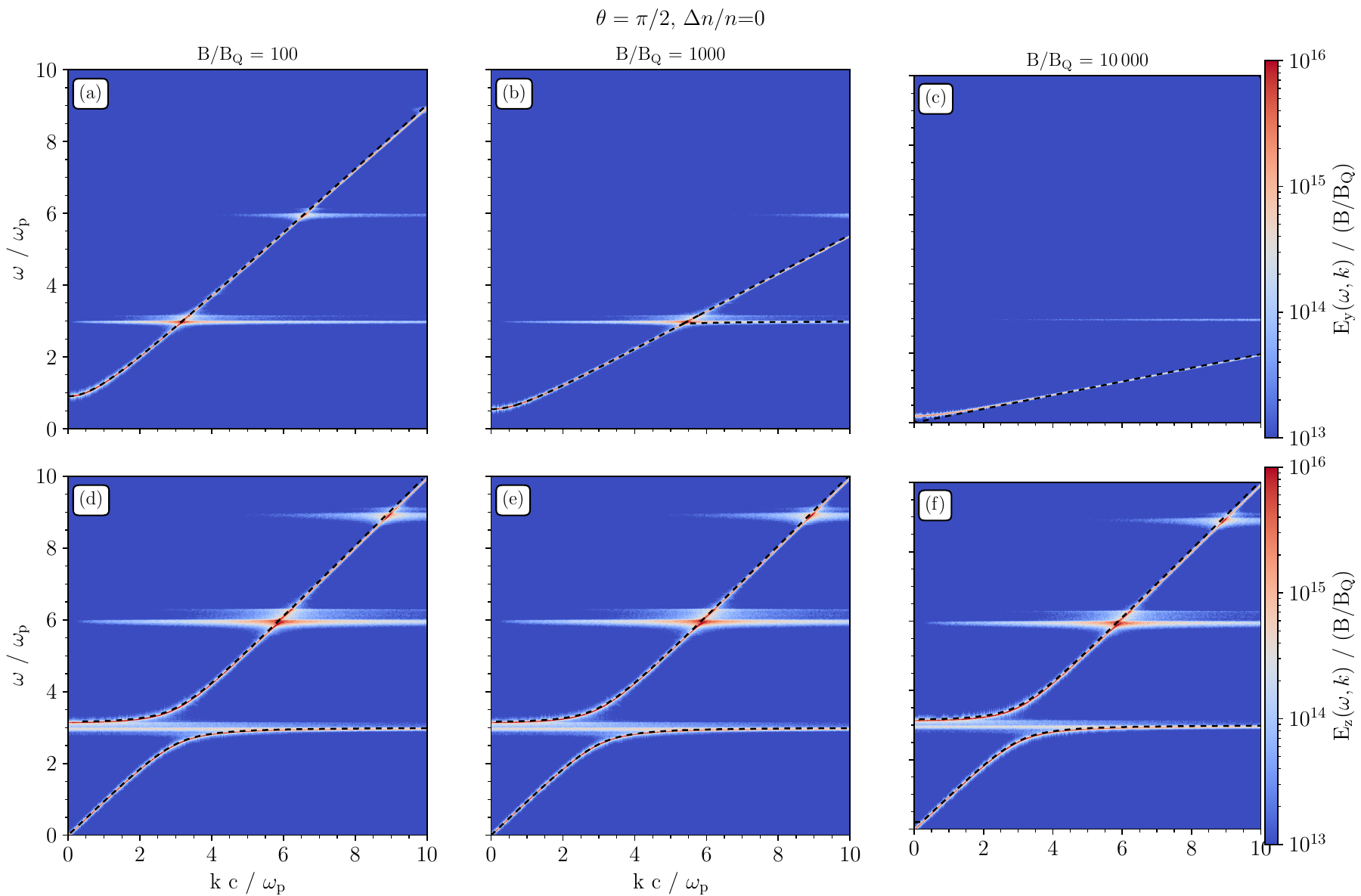}
    \caption{
        The dispersion diagrams of the electric field components on the magnetic field intensity $B/B_\mathrm{Q}$ for $\theta = \pi / 2$ and $\Delta n / n = 0$.
        \label{fig3}
    }
\end{figure}

Figure~\ref{fig3} shows the electric field dispersion properties for increasing magnetic field intensity from $B/B_\mathrm{Q} = 100$ to 10\,000.
The component $E_\mathrm{x}$ is not shown because it does not have significant dependence on $B$ and is similar to figure~\ref{fig1}(a).
While there are no detectable changes in $E_\mathrm{x}$ and $E_\mathrm{z}$ components, the $O$ mode waves in $E_\mathrm{y}$ bend towards lower phase velocities as expected for increasing refractive index $N$ with increasing $\boldsymbol{B}$.

\begin{figure}
    \centering
    \includegraphics[width=\textwidth]{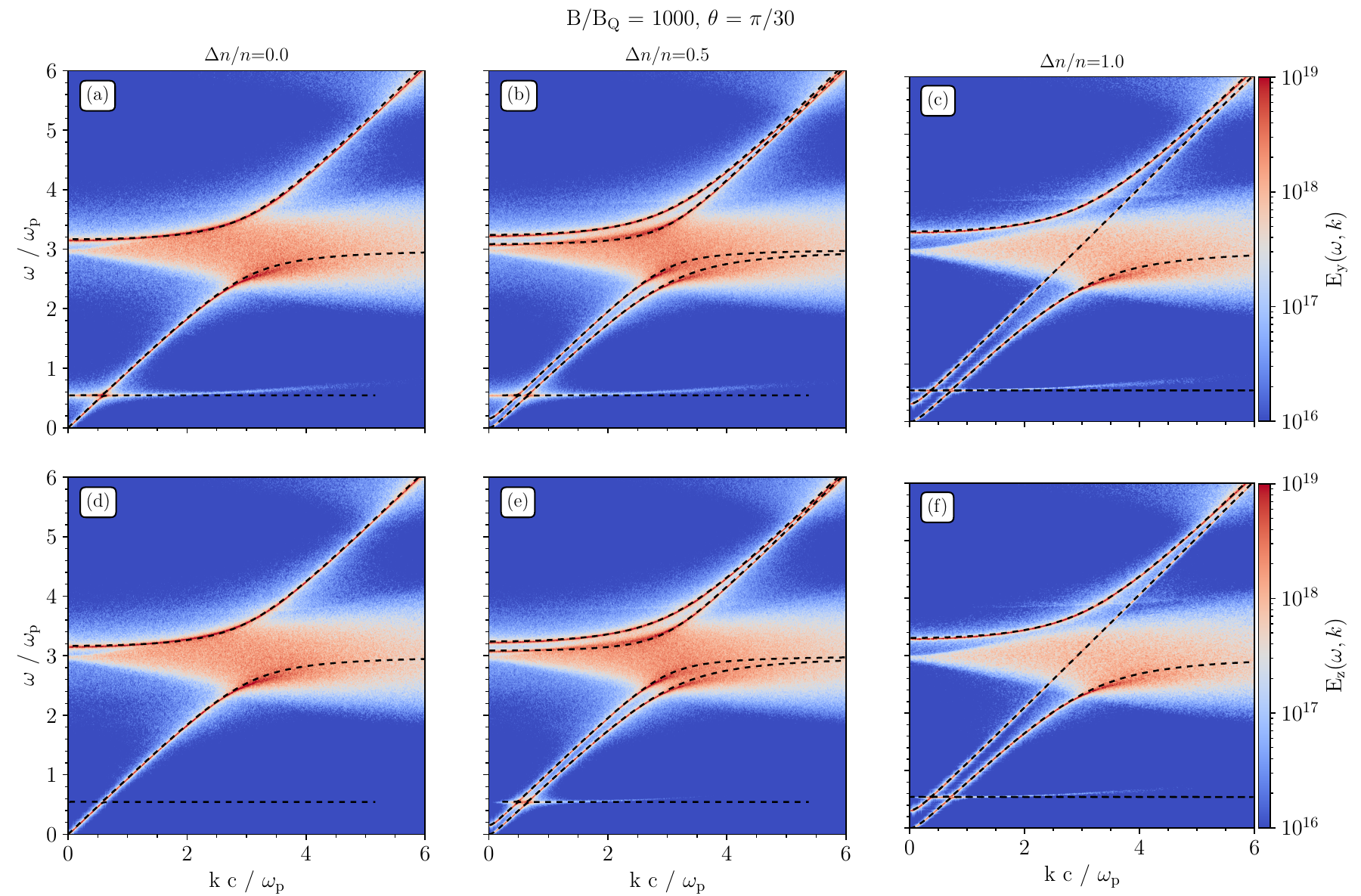}
    \caption{
        The dispersion diagrams of electric field components on the charge density imbalance $\Delta n / n = 0$ for 
        $B/B_\mathrm{Q} = 1000$ and $\theta = \pi / 2$.
        \label{fig4}
    }
\end{figure}

In Figure~\ref{fig4}, we present the dispersion properties of electrostatic waves for $\theta = \pi/30$ with increasing density difference between electrons and positrons.
With the increase, the electromagnetic branches split into two in the frequency, forming two oppositely polarized degeneracies.
Because the electromagnetic waves have circular polarization, the $E_\mathrm{y}$ and $E_\mathrm{z}$ components are associated with the same waves.
Our further analysis showed that the upper mode above $\omega_\mathrm{c}$ is left-handed circularly polarized and the lower is right-handed circularly polarized.
The position of the electrostatic Langmuir mode remains the same in all cases.
The frequency separation increases as the cyclotron frequency.

\subsection{Kinetic effects for ground-level Landau quantization}

\begin{figure}[tb]
    \centering
    \includegraphics[width=\textwidth]{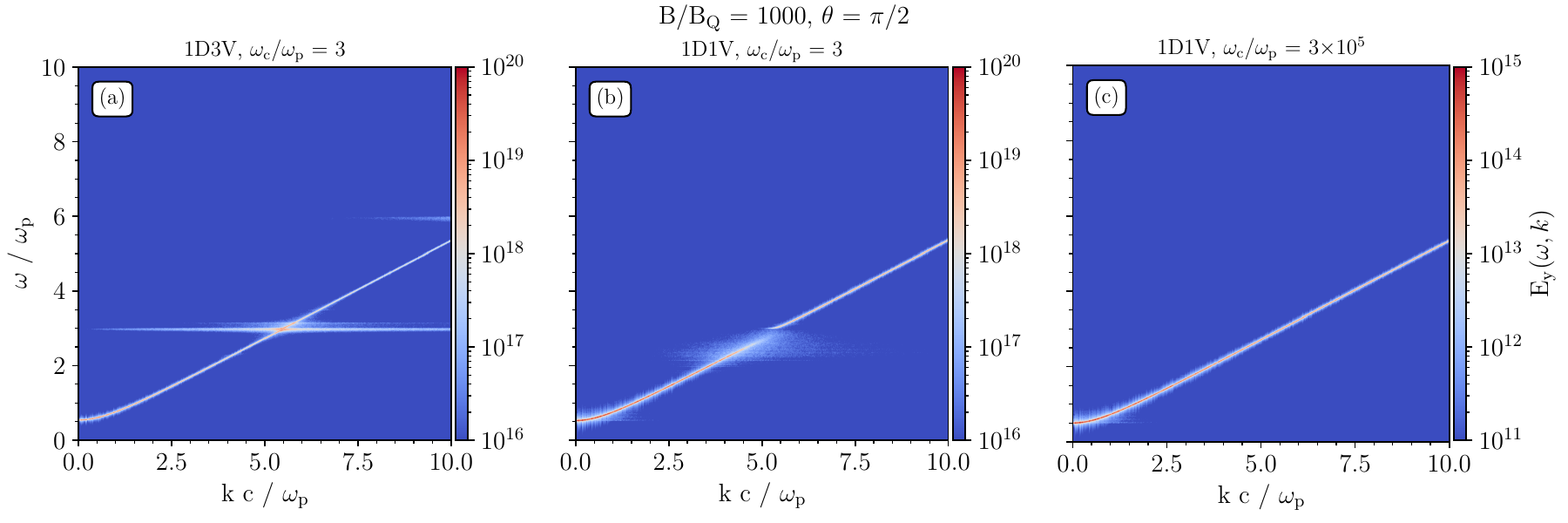}
    \caption{
        Comparison of dispersion properties of the $E_\mathrm{y}$ component for 3-dimensional (1D3V) and 1-dimensional (1D1V) velocity distributions and for plasma frequency decreased by five orders of magnitude, demonstrating the advantage of the QED polarization solver in the case of unresolved cyclotron frequency in the gyro-motion, gyro-center, and gyro-kinetic approximations.
        The other parameters are fixed to $B/B_\mathrm{Q}$, $\theta = \pi / 2$, $\Delta n / n = 0$, and $v_\mathrm{t}/c = 0.05$.
        \label{fig6}
    }
\end{figure}

In the quantum regime, the particle's perpendicular momenta are quantized into Landau levels.
The transitions between the levels are done by absorbing or emitting a quantum of energy.
Though this process is not described in the simulations, we can employ the ground level by setting the perpendicular velocity to zero, corresponding to the case that the perpendicular momenta are radiated by the cyclotron losses.

In addition, the usage of a very small ratio of $\omega_\mathrm{c}/\omega_\mathrm{p} = 3$ for our tests, which cannot be achieved in realistic plasma, 
allow us to study the dispersion properties around both the plasma frequency and the cyclotron frequency.
Nonetheless, the question opens whether the solver also provides reliable outputs when $\omega_\mathrm{p} \lll \omega_\mathrm{c}$ in the magnetar magnetospheres or laser plasmas.

To test also these aspects, we conducted two additional simulations with one-dimensional velocity distribution (1D1V) with a nonzero velocity component along the magnetic field for fixed $B/B_\mathrm{Q}$, $\theta = \pi / 2$, $\Delta n / n = 0$, and $v_\mathrm{t}/c = 0.05$.
In the second of these simulations, we also separated the cyclotron and frequencies by setting the ratio  $\omega_\mathrm{c}/\omega_\mathrm{p} = 3\times10^5$ but adjusting the time step and grid cell sizes to resolve the processes close to the plasma frequency ($\Delta x / d_\mathrm{e} = 0.05$ and $\Delta t \omega_\mathrm{p} = 0.02$).
Thus, the cyclotron frequency and the Larmor radius are not resolved by the time step size and the grid cell size, respectively.

The results are shown in figure~\ref{fig6} and compared with the simulation having three-dimensional velocity distribution (1D3V).
While the cyclotron effects are not well described in Fig.~\ref{fig6}(b) and not present in Fig.~\ref{fig6}(c), the plasma effects as well as the QED polarization effects are present in both cases (b--c).
Therefore, the developed framework can be utilized also in simulations which resolve only the electron plasma and lower frequencies and describe the high-frequency particle Larmor motion by gyro-motion, gyro-center, or gyro-kinetic approximations.

\section{Discussion}
\label{sec:discuss}

Our investigation in this paper focused on reproducing in a PIC simulation the modifications in the dispersion relations of electromagnetic waves that are imposed by QED effects of the super-strong magnetic field exceeding the Schwinger limit
The newly developed solver for QED polarization reproduces birefringence effects between $O$ and $X$ modes, and the dispersion properties follow the analytical solutions derived for cold plasma.

The splitting/degeneracy of $L$ and $R$ modes could cause an observable time delay of circularly polarized waves, if the waves are emitted below or close to the cyclotron frequency.
Such a delay could indicate a nonzero charge density in the emission and propagation regions.

\subsection{Thermal effects in plasma}
One of the main advantages of introducing the electromagnetic-field solver for QED polarization is its ability to be exploited for kinetic plasma simulations. The kinetic simulations, as opposed to nonlinear Maxwell equations alone, can self-consistently describe nonlinear wave–wave and particle–wave effects and also the time evolution of plasma instabilities. In addition, the kinetic simulations can deal with finite-temperature modifications of the dispersion properties.

The thermal effects are one of the reasons for differences between the analytically derived and simulated dispersion relations.
Among other differences is the numerical bending of the light line in the frequency/wave-number space towards lower frequencies at high wave number, $k$, which can potentially cause numerical Cerenkov radiation, if particles with a speed close to the light speed are present in the simulation.
Our results show that the potentially produced numerical Cerenkov radiation in our implementation is comparable with that seen with the classical second-order non-QED field solver.

\subsection{Advantages and limitations of the QED polarization approach}
There are a few effects of very strong magnetic fields that our simulation can not capture. 
Cyclotron radiation is a significant cooling mechanism for electrons in super-strong magnetic fields, and most of the plasma energy can be released as electromagnetic radiation due to the extremely fast gyration of the electrons. For electrons the ratio of the cyclotron cooling time, $\tau_{cyc,cool}$, with the Larmor gyration time, $\tau_{g}$, is
\begin{equation}\label{eq:cyc}
    \frac{\tau_{cyc,cool}}{\tau_{g}} = \frac{9}{16 \pi \alpha}\left(\frac{B_Q}{B}\right),
\end{equation}
where $\alpha$ is the fine-structure constant. Equation \ref{eq:cyc} yields a timescale ratio of 0.25 for a magnetic field a hundred times as strong as the critical field. This implies a substantial energy loss in the test simulations due to the electromagnetic radiation. However, since we have a simulation setup with a periodic boundary condition, the energy in the system is conserved and the emitted electromagnetic radiation eventually interacts with the particles transferring the energy back and forth between the fields and the particles.
Hence, the electromagnetic waves cannot escape the closed simulation system, and no energy is carried away. 
After the simulation starts, an equilibrium between particles and waves is reached and the particle distribution function in the perpendicular direction does not significantly change over the whole simulation time. 


Furthermore, the quantum nature of particles becomes important when the particle separation is smaller or comparable to the De Broglie wavelength, $\lambda_{D}= \frac{h}{m_e v_{th}} \sim 7.8 \times 10^{-10}$ cm, where the particle velocity is taken as the thermal velocity in the test simulations,$v_{th} = 0.05 c$. This condition is satisfied for a density of $2.1 \times 10^{27}$ cm$^{-3}$ or higher. 

Even though this critical density is much higher than the typical plasma densities around the neutron star's surface, $10^{12} - 10^{18}$ cm$^{-3}$, we reach densities in the test simulations for which the wave-like nature of particles is important. The reason is that we require a fixed frequency ratio of $\omega_\text{c}/\omega_\text{p} = 3$ to compare with the analytically derived dispersion curves. For example, for a magnetic strength of 10$^4$\,$B_Q$, the implied plasma density is around $5\times 10^{39}$cm$^{-3}$, which is far higher than the critical density of $2.1 \times 10^{27}$ cm$^{-3}$. In principle, one should include the particle quantization in the simulation setup. However, the goal of our test simulations is to test the QED polarization field solver which is independent of these quantization effects.

The newly developed numerical field solver for QED polarization is one ingredient of a fully self-consistent kinetic model of plasma in super-strong magnetic field. Another effect not included in our approach is the photon annihilation with the magnetic field which produces new pairs. 



\section{Conclusions}
\label{sec:conclude}
To model magnetic fields exceeding the Schwinger limit we developed a field solver including QED polarization and implemented it in a PIC plasma code. With this extension, the PIC code allows us to investigate super-strongly magnetized plasma environments at kinetic scales, as found in the magnetospheres of magnetars or in laboratory plasma experiments with lasers. The advantages of the solver are (1) the applicability for gyro-motion, gyro-center, and gyro-kinetic simulations, which do not resolve the cyclotron motion, (2) the possibility to study with the PIC method plasmas with relativistic temperatures present in, e.g., magnetar magnetospheres and laser plasmas, and (3) to describe the nonlinear evolution of systems harboring time- and spatially-dependent electromagnetic fields in vacuum and plasmas.

The field solver for QED polarization well reproduces the dispersion properties of electrostatic and electromagnetic waves in cold plasma that were derived in \cite{Medvedev2023}.
The $O$ and $X$ mode polarizations show birefringence, which increases with increasing magnetic-field intensity and could explain the polarization properties of high-energy radiation observed from magnetars.
Also, the parallel-propagating electromagnetic mode splits into two modes when the charge density, $\Delta n / n$, becomes nonzero.


In future work, we would like to implement QED effects in the numerical plasma macroparticles, and utilize the code for studying electromagnetic instabilities in magnetar magnetospheres.

\begin{acknowledgments}
We thank Andrew Taylor for the helpful discussion and comments.
We acknowledge the developers of the ACRONYM code (Verein zur Förderung kinetischer Plasmasimulationen e.V.). 
We acknowledge the support by the German Science Foundation (DFG) project BE 7886/2-1.
MM acknowledges support by NSF via grant PHY-2409249.
The authors gratefully acknowledge the Gauss Centre for Supercomputing e.V. (www.gauss-centre.eu) for partially funding this project by providing computing time on the GCS Supercomputer SuperMUC-NG at Leibniz Supercomputing Centre (www.lrz.de), project pn72ku. 
\end{acknowledgments}

\section*{Data Availability Statement}
The data that support the findings of this study are available from the corresponding authors upon reasonable request. 

\appendix
\section{Field solver for QED polarization}\label{app:EMsolver}

Here we derive the explicit electric fields' time derivatives for the modified Ampère's law in the super-strong magnetic fields approach (equation~\eqref{eq:ModAmpere}). We start by expressing the QED parameters ($C_\delta$, $C_\mu$ and $C_\epsilon$) as given by equations~\eqref{eq:C_delta}--\eqref{eq:C_epsilon}. Then we rearrange the there equations of the spatial components of equation~\eqref{eq:ModAmpere} in the following matrix form 

\begin{equation}\label{eq:Matrixform}
    \begin{bmatrix}
    A_\mathrm{xx} & A_\mathrm{xy} & A_\mathrm{xz} \\
    A_\mathrm{yx} & A_\mathrm{yy} & A_\mathrm{yz} \\
    A_\mathrm{zx} & A_\mathrm{zy} & A_\mathrm{zz}
    \end{bmatrix} 
    \begin{bmatrix}
    \frac{\partial E_\mathrm{x}}{\partial t} \\
    \frac{\partial E_\mathrm{y}}{\partial t} \\
    \frac{\partial E_\mathrm{z}}{\partial t}
    \end{bmatrix} = 
    \begin{bmatrix}
    \frac{1}{c}j_\mathrm{x} - Q_\mathrm{x} \\
    \frac{1}{c}j_\mathrm{y} - Q_\mathrm{y} \\
    \frac{1}{c}j_\mathrm{z} - Q_\mathrm{z}
    \end{bmatrix},
\end{equation}
where the components of the matrix $A$ are given by
\begin{equation}
    A_{ij} = \frac{1}{c} \left[ \gamma_{\mathcal{F}} \delta_{ij} - \gamma_{\mathcal{F} \mathcal{F}} E_i E_j - \gamma_{\mathcal{G} \mathcal{G}} B_i B_j \right],
\end{equation}
with $i$ and $j$ looping over the spatial coordinates $x$, $y$ and $z$. Here $j_x$, $j_y$ and $j_z$ are the spatial components of the electric current. The additional $Q$ terms on the right-hand side include all the spatial derivatives of the electric and magnetic fields along with the time derivatives of the magnetic fields and given as

\begin{equation}\label{eq:Qx}
    \begin{split}
        Q_\mathrm{x} = & \gamma_{\mathcal{F}} \left[-\left(\frac{\partial B_\mathrm{z}}{\partial y} - \frac{\partial B_\mathrm{y}}{\partial z} \right)\right] \\ & + \gamma_{\mathcal{F} \mathcal{F}}\left[ \frac{1}{c} E_\mathrm{x} \left(\vb{B}\cdot\frac{\partial \vb{B}}{\partial t}\right) + B_\mathrm{y} \left( \vb{B}\cdot\frac{\partial \vb{B}}{\partial z} - \vb{E}\cdot\frac{\partial \vb{E}}{\partial z}\right) - B_\mathrm{z} \left( \vb{B}\cdot\frac{\partial \vb{B}}{\partial y} - \vb{E}\cdot\frac{\partial \vb{E}}{\partial y}\right)\right] \\ & + \gamma_{\mathcal{G} \mathcal{G}} \left[ - \frac{1}{c} B_\mathrm{x} \left( \vb{E}\cdot\frac{\partial \vb{B}}{\partial t} \right) + E_\mathrm{y} \left( \vb{E}\cdot\frac{\partial \vb{B}}{\partial z} + \vb{B}\cdot\frac{\partial \vb{E}}{\partial z}\right) - E_\mathrm{z} \left( \vb{E}\cdot\frac{\partial \vb{B}}{\partial y} - \vb{B}\cdot\frac{\partial \vb{E}}{\partial y}\right)\right],
    \end{split}
\end{equation}

\begin{equation}\label{eq:Qy}
    \begin{split}
        Q_\mathrm{y} = & \gamma_{\mathcal{F}} \left[-\left(\frac{\partial B_\mathrm{x}}{\partial z} - \frac{\partial B_\mathrm{z}}{\partial x} \right)\right] \\ & + \gamma_{\mathcal{F} \mathcal{F}}\left[ \frac{1}{c} E_\mathrm{y} \left(\vb{B}\cdot\frac{\partial \vb{B}}{\partial t}\right) + B_\mathrm{z} \left( \vb{B}\cdot\frac{\partial \vb{B}}{\partial x} - \vb{E}\cdot\frac{\partial \vb{E}}{\partial x}\right) - B_\mathrm{x} \left( \vb{B}\cdot\frac{\partial \vb{B}}{\partial z} - \vb{E}\cdot\frac{\partial \vb{E}}{\partial z}\right)\right] \\ & + \gamma_{\mathcal{G} \mathcal{G}} \left[ - \frac{1}{c} B_\mathrm{y} \left( \vb{E}\cdot\frac{\partial \vb{B}}{\partial t} \right) + E_\mathrm{z} \left( \vb{E}\cdot\frac{\partial \vb{B}}{\partial x} + \vb{B}\cdot\frac{\partial \vb{E}}{\partial x}\right) - E_\mathrm{x} \left( \vb{E}\cdot\frac{\partial \vb{B}}{\partial z} - \vb{B}\cdot\frac{\partial \vb{E}}{\partial z}\right)\right],
    \end{split}
\end{equation}
and
\begin{equation}\label{eq:Qz}
    \begin{split}
        Q_\mathrm{z} = & \gamma_{\mathcal{F}} \left[-\left(\frac{\partial B_\mathrm{y}}{\partial x} - \frac{\partial B_\mathrm{x}}{\partial y} \right)\right] \\ & + \gamma_{\mathcal{F} \mathcal{F}} \left[ \frac{1}{c} E_\mathrm{z} \left(\vb{B}\cdot\frac{\partial \vb{B}}{\partial t}\right) + B_\mathrm{x} \left( \vb{B}\cdot\frac{\partial \vb{B}}{\partial y} - \vb{E}\cdot\frac{\partial \vb{E}}{\partial y}\right) - B_\mathrm{y} \left( \vb{B}\cdot\frac{\partial \vb{B}}{\partial x} - \vb{E}\cdot\frac{\partial \vb{E}}{\partial x}\right)\right] \\ & + \gamma_{\mathcal{G} \mathcal{G}} \left[ - \frac{1}{c} B_\mathrm{z} \left( \vb{E}\cdot\frac{\partial \vb{B}}{\partial t} \right) + E_\mathrm{x} \left( \vb{E}\cdot\frac{\partial \vb{B}}{\partial y} + \vb{B}\cdot\frac{\partial \vb{E}}{\partial y}\right) - E_\mathrm{y} \left( \vb{E}\cdot\frac{\partial \vb{B}}{\partial x} - \vb{B}\cdot\frac{\partial \vb{E}}{\partial x}\right)\right].
    \end{split}
\end{equation}

Finally, multiplying equation~\eqref{eq:Matrixform} with $A^{-1}$ we get the explicit form of the electric field temporal evolution

\begin{equation}\label{eq:solver}
    \frac{\partial \vb{E}}{\partial t} = A^{-1} \left( \frac{1}{c} \vb{j} - \vb{Q}\right),
\end{equation}
where
\begin{equation}
    \begin{split}
        A^{-1} = \frac{1}{|A|}     \begin{bmatrix}
    A_\mathrm{yy}A_\mathrm{zz}-A_\mathrm{zy}A_\mathrm{yz} & A_\mathrm{xz}A_\mathrm{zy}-A_\mathrm{zz}A_\mathrm{xy} & A_\mathrm{xy}A_\mathrm{yz}-A_\mathrm{yy}A_\mathrm{xz} \\
    A_\mathrm{yz}A_\mathrm{zx}-A_\mathrm{zz}A_\mathrm{yx} & A_\mathrm{xx}A_\mathrm{zz}-A_\mathrm{zx}A_\mathrm{xz} & A_\mathrm{xz}A_\mathrm{yx}-A_\mathrm{yz}A_\mathrm{xx} \\
    A_\mathrm{yx}A_\mathrm{zy}-A_\mathrm{zx}A_\mathrm{yy} & A_\mathrm{xy}A_\mathrm{zx}-A_\mathrm{zy}A_\mathrm{xx} & A_\mathrm{xx}A_\mathrm{yy}-A_\mathrm{yx}A_\mathrm{xy}
    \end{bmatrix},
    \end{split}
\end{equation}
and
\begin{equation}
    |A| = A_\mathrm{xx} \left( A_\mathrm{yy}A_\mathrm{zz}-A_\mathrm{zy}A_\mathrm{yz} \right) - A_\mathrm{xy} \left( A_\mathrm{yx}A_\mathrm{zz}-A_\mathrm{zx}A_\mathrm{yz} \right) + A_\mathrm{xz} \left( A_\mathrm{yx}A_\mathrm{zy}-A_\mathrm{zx}A_\mathrm{yy} \right).
\end{equation}




\subsection{Implementation in 1D3V PIC}

When implementing the QED polarization solver (equation~\eqref{eq:solver}) in a 1D3V PIC code (one spatial dimension, three velocity and electromagnetic field components), the only part that is changed is the $\vb{Q}$ vector. If we have the simulation spatial extend in the $x$-axis, then we get the following expressions for the components of the $\vb{Q}$ vector

\begin{equation}\label{eq:Qx1D}
    \begin{split}
        Q_\mathrm{x} = \gamma_{\mathcal{F} \mathcal{F}}\left[ \frac{1}{c} E_\mathrm{x} \left(\vb{B}\cdot\frac{\partial \vb{B}}{\partial t}\right) \right]  + \gamma_{\mathcal{G} \mathcal{G}} \left[ - \frac{1}{c} B_\mathrm{x} \left( \vb{E}\cdot\frac{\partial \vb{B}}{\partial t} \right) \right],
    \end{split}
\end{equation}
\begin{equation}\label{eq:Qy1D}
    \begin{split}
        Q_\mathrm{y} = & \gamma_{\mathcal{F}} \left[\left(\frac{\partial B_\mathrm{z}}{\partial x} \right)\right] \\ & + \gamma_{\mathcal{F} \mathcal{F}}\left[ \frac{1}{c} E_\mathrm{y} \left(\vb{B}\cdot\frac{\partial \vb{B}}{\partial t}\right) + B_\mathrm{z} \left( \vb{B}\cdot\frac{\partial \vb{B}}{\partial x} - \vb{E}\cdot\frac{\partial \vb{E}}{\partial x}\right)  \right] \\ & + \gamma_{\mathcal{G} \mathcal{G}} \left[ - \frac{1}{c} B_\mathrm{y} \left( \vb{E}\cdot\frac{\partial \vb{B}}{\partial t} \right) + E_\mathrm{z} \left( \vb{E}\cdot\frac{\partial \vb{B}}{\partial x} + \vb{B}\cdot\frac{\partial \vb{E}}{\partial x}\right)  \right],
    \end{split}
\end{equation}
and
\begin{equation}\label{eq:Qz1D}
    \begin{split}
        Q_\mathrm{z} = & \gamma_{\mathcal{F}} \left[-\left(\frac{\partial B_\mathrm{y}}{\partial x}  \right)\right] \\ & + \gamma_{\mathcal{F} \mathcal{F}}\left[ \frac{1}{c} E_\mathrm{z} \left(\vb{B}\cdot\frac{\partial \vb{B}}{\partial t}\right) - B_\mathrm{y} \left( \vb{B}\cdot\frac{\partial \vb{B}}{\partial x} - \vb{E}\cdot\frac{\partial \vb{E}}{\partial x}\right)\right] \\ & + \gamma_{\mathcal{G} \mathcal{G}} \left[ - \frac{1}{c} B_\mathrm{z} \left( \vb{E}\cdot\frac{\partial \vb{B}}{\partial t} \right) - E_\mathrm{y} \left( \vb{E}\cdot\frac{\partial \vb{B}}{\partial x} - \vb{B}\cdot\frac{\partial \vb{E}}{\partial x}\right)\right].
    \end{split}
\end{equation}
For the magnetic field time derivative in the calculations of the $\vb{Q}$ components (equations~\eqref{eq:Qx1D}--\eqref{eq:Qz1D}) we substitute its value from Faraday's law
\begin{equation}
    \frac{\partial \vb{B}}{\partial t} = - c \curl \vb{E} = (0, c \frac{\partial E_\mathrm{z}}{\partial x}, -c \frac{\partial E_\mathrm{y}}{\partial x}).
\end{equation}
Arriving at expressions for the $\vb{Q}$ vector components dependent only on the electromagnetic fields' spatial derivatives.

\section{Numerical stability of nonlinear Maxwell equations} \label{app:NS}


Here, we drive the numerical dispersion relation of a wave propagating on a one-dimensional Yee grid following the nonlinear Maxwell equations. Assuming a homogenise super-strong magnetic field in the (x,y) plane, $\mathbf{B_0} = B_0 \cos{\theta} \hat{x} + B_0 \sin{\theta} \hat{y}$ and wave-like perturbations in electric and magnetic fields, we find the numerical dispersion relation for two different electromagnetic wave modes ($X$ and $O$ modes). 


For the X mode, we have the following wave-like perturbation components
\begin{equation}
    \delta B_\mathrm{y} = \delta B_\mathrm{y0} \exp{i(\omega \Delta t - k_\mathrm{x} \Delta x)},
\end{equation}
\begin{equation}
    \delta E_\mathrm{x} = \delta E_\mathrm{x0} \exp{i(\omega \Delta t - k_\mathrm{x} \Delta x)},
\end{equation}
and
\begin{equation}
    \delta E_\mathrm{z} = \delta E_\mathrm{z0} \exp{i(\omega \Delta t - k_\mathrm{x} \Delta x)}.
\end{equation}

Using the Faraday law (equation~\eqref{eq:Faraday}), we get the following relation between $\delta B_\mathrm{y}$ and $\delta E_\mathrm{z}$ from the $y$ spatial component
\begin{equation} \label{eq:By0}
    \delta B_\mathrm{y} = - \left( \frac{c \Delta t}{\Delta x} \right) \frac{\sin\left( \frac{k \Delta x}{2} \right)}{\sin\left( \frac{\omega \Delta t}{2} \right)} \delta E_\mathrm{z}.
\end{equation}

Using the QED-modified Ampère's law (equation~\eqref{eq:ModAmpere}), we get from the $z$ component the following relation
\begin{equation}
\begin{split}
    \Bigg[ & \left( \frac{1}{c} i \frac{\sin \left(\frac{\omega \Delta t}{2}\right)}{\Delta t /2} \delta E_\mathrm{z} + i \frac{\sin \left(\frac{k_\mathrm{x} \Delta x}{2}\right)}{\Delta x /2} \delta B_\mathrm{y} \right) \left( \gamma_\mathcal{F} +\gamma_{\mathcal{FF}} \left(\delta\mathbf{B}^2 +\delta\mathbf{B} \cdot \mathbf{B_0} - \delta\mathbf{E}^2 \right)  \right) \\ & + \left( i \frac{\sin \left(\frac{k_\mathrm{x} \Delta x}{2}\right)}{\Delta x /2} B_\mathrm{y0} \right) \left(\gamma_{\mathcal{FF}} \left(\delta\mathbf{B}^2 +\delta\mathbf{B} \cdot \mathbf{B_0} - \delta\mathbf{E}^2 \right)  \right) \\ & + \gamma_{\mathcal{GG}}  \left( \frac{1}{c} i \frac{\sin \left(\frac{\omega \Delta t}{2}\right)}{\Delta t /2} \delta B_\mathrm{z} - i \frac{\sin \left(\frac{k_\mathrm{x} \Delta x}{2}\right)}{\Delta x /2} \delta E_\mathrm{y}  \right) \left(-\mathbf{B_0} \cdot \delta\mathbf{E} -2 \delta\mathbf{B} \cdot \delta\mathbf{E} \right)  \\ & + \gamma_{\mathcal{GG}}  \left( \frac{1}{c} i \frac{\sin \left(\frac{\omega \Delta t}{2}\right)}{\Delta t /2} B_\mathrm{z0}  \right) \left(-\mathbf{B_0} \cdot \delta\mathbf{E} -2 \delta\mathbf{B} \cdot \delta\mathbf{E} \right)  \Bigg] = 0,
\end{split}
\end{equation}
where $\gamma_\mathcal{F} = - (1-C_\delta)/(4\pi)$, $\gamma_{\mathcal{FF}} = C_\mu/(4\pi B^2)$, and $\gamma_{\mathcal{GG}} = C_\epsilon/(4\pi B^2)$. Assuming that the wave-like perturbations always stay much smaller than the homogenies super-strong magnetic field, $\delta B, \delta E \ll B_0$, then we neglect the terms of the orders $\delta B/B_0$, $\delta E/B_0$, $(\delta B/B_0)^2$ and $(\delta E/B_0)^2$, and keep the terms of the order of $\delta B$ and $\delta E$ arriving at the following relation
\begin{equation}
\begin{split}
    \Bigg[ & \gamma_\mathcal{F} \left( \frac{1}{c} \frac{\sin \left(\frac{\omega \Delta t}{2}\right)}{\Delta t} \delta E_\mathrm{z} +  \frac{\sin \left(\frac{k_\mathrm{x} \Delta x}{2}\right)}{\Delta x} \delta B_\mathrm{y} \right) \\ & + \gamma_{\mathcal{FF}} \frac{\sin \left(\frac{k_\mathrm{x} \Delta x}{2}\right)}{\Delta x} B_\mathrm{y0}  \left(\delta\mathbf{B} \cdot \mathbf{B_0}\right) \\ & - \gamma_{\mathcal{GG}}  \frac{1}{c} \frac{\sin \left(\frac{\omega \Delta t}{2}\right)}{\Delta t} B_{z0}  \left(\mathbf{B_0} \cdot \delta\mathbf{E}\right) \Bigg] = 0.
\end{split}
\end{equation}
Substituting the perturbation components for the X-mode we get
\begin{equation}\label{eq:TMz0}
\begin{split}
& \gamma_\mathcal{F} \left( \frac{1}{c} \frac{\sin \left(\frac{\omega \Delta t}{2}\right)}{\Delta t} \delta E_\mathrm{z0} +  \frac{\sin \left(\frac{k_\mathrm{x} \Delta x}{2}\right)}{\Delta x} \delta B_\mathrm{y0} \right) \\ & + \gamma_{\mathcal{FF}} \frac{\sin \left(\frac{k_\mathrm{x} \Delta x}{2}\right)}{\Delta x} B_0^2 \sin^2{\theta} \delta B_\mathrm{y} = 0,
\end{split}
\end{equation}
where $\theta$ is the angle of the magnetic field with the wave propagation direction (fixed to the $x$ axis). Substituting equation \ref{eq:By0} in equation \ref{eq:TMz0}, we finally find the dispersion relation for the X mode waves
\begin{equation}\label{eq:Xmode}
    \left( \frac{1-C_\delta - C_\mu \sin^2{\theta}}{1-C_\delta} \right)\left( \frac{c \Delta t}{\Delta x} \right)^2 \sin^2 \left( \frac{k \Delta x}{2} \right) - \sin^2 \left( \frac{\omega \Delta t}{2} \right) = 0.
\end{equation}



Now we go back to the O-mode waves, which have the following wave-like components 
\begin{equation}
    \delta B_\mathrm{z} = \delta B_\mathrm{z0} \exp{i(\omega \Delta t - k_\mathrm{x} \Delta x)},
\end{equation}
\begin{equation}
    \delta E_\mathrm{x} = \delta E_\mathrm{x0} \exp{i(\omega \Delta t - k_\mathrm{x} \Delta x)},
\end{equation}
\begin{equation}
    \delta E_\mathrm{y} = \delta E_\mathrm{y0} \exp{i(\omega \Delta t - k_\mathrm{x} \Delta x)}.
\end{equation}

Using the QED-modified Ampère's law (equation~\eqref{eq:ModAmpere}), we get from the $y$ component the following relation 
\begin{equation} \label{eq:TEz0}
\begin{split}
 & \gamma_F \left( \frac{1}{c} \frac{\sin \left(\frac{\omega \Delta t}{2}\right)}{\Delta t} \delta E_\mathrm{y} -  \frac{\sin \left(\frac{k_\mathrm{x} \Delta x}{2}\right)}{\Delta x} \delta B_\mathrm{z} \right) \\ &  - \gamma_\mathrm{gg}  \frac{1}{c} \frac{\sin \left(\frac{\omega \Delta t}{2}\right)}{\Delta t} B_0 \sin{\theta}  \left( B_0 \cos{\theta} \delta E_\mathrm{x} + B_0 \sin{\theta} \delta E_\mathrm{y} \right) = 0,
\end{split}
\end{equation}
and the following relation from the $x$ component 
\begin{equation}\label{eq:Ex0}
    \delta E_\mathrm{x} = \frac{\gamma_\mathrm{gg}B_0^2 \cos{\theta} \sin{\theta}}{\gamma_F - \gamma_\mathrm{gg} B_0^2 \cos^2{\theta}} \delta E_\mathrm{y}. 
\end{equation}
We also get a third relation from the $z$ of the Faraday's law (equation \ref{eq:Faraday})
\begin{equation}\label{eq:Bz0}
    \delta B_\mathrm{z} = \left( \frac{c \Delta t}{\Delta x} \right) \frac{\sin\left( \frac{k \Delta x}{2} \right)}{\sin\left( \frac{\omega \Delta t}{2} \right)} \delta E_\mathrm{y}.
\end{equation}

Finally after substituting the equations~\eqref{eq:Ex0} and \eqref{eq:Bz0}, in equation~\eqref{eq:TEz0}, we get the dispersion relation for the $O$ mode waves that is
\begin{equation}\label{eq:Omode}
    \left( \frac{1-C_\delta + C_\epsilon \cos^2{\theta}}{1-C_\delta + C_\epsilon} \right)\left( \frac{c \Delta t}{\Delta x} \right)^2 \sin^2 \left( \frac{k \Delta x}{2} \right) - \sin^2 \left( \frac{\omega \Delta t}{2} \right) = 0.
\end{equation}


\nocite{*}
\bibliography{main}

\end{document}